\documentclass{aa}  
\usepackage{amsmath}
\usepackage{tablefootnote}
\usepackage{bm}
\usepackage{xcolor}
\usepackage{graphicx}
\usepackage{txfonts}
\usepackage[colorlinks,linkcolor=blue,citecolor=blue,linktocpage=true,breaklinks, 
plainpages=false,urlcolor=blue]{hyperref}

\begin{document}

   \title{Mimicking spectropolarimetric inversions using convolutional neural networks}

   \author{I. Mili\'{c}
          \inst{1,2,3}
          \and
          R. Gafeira
          \inst{4,5}
          }

   \institute{Department of Physics, University of Colorado, Boulder CO 80309, USA\\
   \email{ivan.milic@colorado.edu}
   \and
   Laboratory for Atmospheric and Space Physics, University of Colorado, Boulder CO 80303, USA\\
   \and
   National Solar Observatory, Boulder CO 80303, USA\\
   \and
   Instituto Astrofisica de Andalucia (CSIC), Apartado de Correos 3004, E-18080 Granada, Spain \\
   \and 
   Centre for Earth and Space Research of University of Coimbra, Department of Physics, University of Coimbra, 3000-456 Coimbra, Portugal\\
   \email: gafeira@iaa.es
   }

   \date{Received ; accepted }

% \abstract{}{}{}{}{} 
% 5 {} token are mandatory
 
  \abstract
   {Interpreting spectropolarimetric observations of the solar atmosphere takes much  longer  than the acquiring the data. The most important reason for this is that the model fitting, or ``inversion'', used to infer physical quantities from the observations is extremely slow, because the underlying models are numerically demanding.} 
   {We aim to improve the speed of the inference by using a neural network that relates input polarized spectra to the output physical parameters.}
   {We first select a subset of the data to be interpreted and infer physical quantities from corresponding spectra using a standard minimization-based inversion code. Taking these results as reliable and representative of the whole  data set, we train a convolutional neural network to connect the input polarized spectra to the output physical parameters (nodes, in context of spectropolarimetric inversion). We then apply the neural network to the various other data, previously unseen to the network. As a check, we apply the referent inversion code to the unseen data and compare the fit quality and the maps of the inferred parameters between the two inversions.}
   {The physical parameters inferred by the neural network show excellent agreement with the results from the inversion, and are obtained in a factor of $10^5$ less time. Additionally, substituting the results of the neural network back in the forward model, shows excellent agreement between inferred and original spectra.}
   {The method we present here is very simple for implementation and extremely fast. It only requires a training data set, which can be obtained by inverting a representative subset of the observed data. Applying these (and similar) machine learning techniques will yield orders of magnitude acceleration in the routine interpretation of spectropolarimetric data.}


   \maketitle
   
%
%________________________________________________________________

\section{Introduction}

Spatially resolved observations of optically thick spectral lines carry a wealth of information about the solar atmosphere. In particular, the spectral lines are formed over a range of depths, and thus carry the information about depth dependence of physical parameters  (e.g., temperature, velocity, magnetic field) in the solar atmosphere. High spatial and spectral resolution observations then allow us to perform  3D mapping of the solar atmosphere. The method of choice for the interpretation of the polarized spectra is via   spectropolarimetric inversions. These are numerical methods that fit a model atmosphere to the observed polarized spectra. Inversions are notoriously numerically demanding;  even the calculation of the forward problem (i.e., evaluation of the spectra of a guess model atmosphere) involves complicated physics of the radiative transfer that needs to be solved numerically.

Initially, a way to handle this was to use the  Milne-Eddington approximation \citep{Auer77,Egidio82}. This, severely simplified model atmosphere, allows an analytical solution of the radiative transfer equation, which means that the model atmosphere can be fit to the observations extremely rapidly. Some of the inversion codes based on the Milne-Eddington approximation are still routinely used to interpret spectropolarimetric observations of the Sun \citep[e.g.,][]{HAO1987,MILOS, VISF2011}. The main issue with this approach is that the magnetic field and line-of-sight velocity are assumed to be constant with depth and the model does not include the temperature. A big step forward was the SIR code by \citet{SIR} that introduced stratified atmospheric models with depth-dependent temperature, pressure, line-of-sight velocity, microturbulence, and the magnetic field vector. The Stokes spectra could then be inverted (i.e., fit) to yield fully stratified model atmospheres. The inversion of a Stokes spectra map, pixel-by-pixel, thus results in a 3D model of the observed region on solar surface. There are  two main limitations of SIR \citep[and similar codes, such as SPINOR by][]{SPINOR}: i) the assumption of local thermodynamic equilibrium (LTE), which makes the codes inappropriate for lines formed by scattering (e.g., Na\,I\,D, Mg\,I\,b, H\,$\alpha$, Ca\,II\,8542, Mg\,II h\&k); ii) the assumption that the atmosphere is in hydrostatic equilibrium. This is necessary in order to eliminate pressure as a free parameter, mostly because it is otherwise too degenerate with the temperature. However, there are no good reasons to assume that the atmosphere is indeed in the hydrostatic equilibrium, especially when we consider spectral lines sampling the upper photosphere and the chromosphere.

A lot has been done to go beyond the LTE approximation and implementation of the so-called nonlocal thermodynamic equilibrium (NLTE) effects in the inversion codes. The first  was NICOLE by \citet{Hector1998,NICOLE}, followed by STiC \citep{STIC} and SNAPI \citep{SNAPI}. There are implementation differences between these codes (especially that STiC can deal with the spectral lines formed in the so-called partial frequency redistribution, PRD), but in essence all three retrieve stratified atmosphere models by fitting the observed Stokes spectrum that can include NLTE lines. We point the interested reader to the original publications describing these codes in detail. 

On the contrary, there were not many attempts to avoid hydrostatic equilibrium approximation and infer the pressure stratification together with the temperature one from the observed Stokes spectra. Notable is the work of \citet{MASI}, who proposed a the MHD-assisted inversion technique that matches columns from MHD simulations to the observed spectra in order to reconstruct the observed atmosphere. The inferred 3D model of the atmosphere is then used as an initial model for MHD simulation and the process is supposed to be iterated until convergence. Needless to say, this approach is extremely numerically demanding.

The more complicated the physics of the line formation, the more demanding the inversion code. NLTE inversion codes are particularly time consuming because   the forward evaluation (i.e., spectral synthesis) needs many iterations to get self-consistent solutions of statistical equilibrium and radiative transfer equations \citep[see, e.g., monograph by][for details on numerical solutions of NLTE problem]{SAbook2014}. Even in the LTE case, the inversion time per pixel is on the order of seconds, and for NLTE order(s) of magnitude more. The next generation of instruments, such as VTF at the soon-to-be-operational DKIST telescope \citep{DKIST}, will provide us with with tens of millions of spectra per observation. For more complicated spectral lines this easily translates into tens of CPU years. This is not only time demanding, but also expensive  in terms of funds and of the ecological fingerprint. Therefore, it is in our best interest to develop inversion methods that will allow us to interpret the data faster.

Particularly interesting in this context are various machine learning (ML) methods, most notably convolutional neural networks \citep[CNNs; ][]{lecun-98,Simard03,CNNref}. In general, neural networks can be understood as pipelines that apply series of very simple transformations on the input (in our case Stokes spectra), finally resulting in the output (in our case atmospheric parameters). Given a training set, the coefficients of the transformations are tuned until the best agreement is obtained between the given output parameters and the ones obtained by applying the neural network to the input. One transformation in this pipeline is called a ``layer.'' Convolutional neural networks are interesting because they involve convolutional layers, that are particularly suitable for feature recognition and characterization. We are interested in applying them to the solar spectra since spectral lines can be viewed as features and their properties are related to the properties of the atmosphere they are generated in.

Neural networks recently started gaining popularity in solar physics. They are used for the inference of plane-of-the-sky velocities, deconvolution and super-resolution, image restoration, and so on. For our research the most relevant are the recent works of \citet{RADYNINV} and \citet{CNN_INV_AA}. Both of these works rely on atmosphere models that are result of the state-of-the art simulations. \citet{RADYNINV} simulate solar flares using the RADYN code \citep{RADYN, RADYN2}, synthesize spectra of the diagnostically interesting lines, and then train an invertible neural network to infer the model atmosphere from the observations. \citet{CNN_INV_AA} use MURAM \citep{MURAM} simulations of sunspots and their quiet Sun surroundings to synthesize spectra of Fe\,I\,6300 lines and then apply a realistic spatial point spread function (PSF) of the HINODE space telescope to their data to simulate the effect of finite spatial resolution. They then train a convolutional neural network on patches of the observations and use it to interpret actual HINODE observations. A network trained in this way automatically performs spatial deconvolution of the data and results in  atmospheres that are spatially smooth. Both of these recent results are, without doubt, a huge leap forward. Inference is orders of magnitude faster than  the case of standard inversions. Additionally, \citet{CNN_INV_AA} are also able to infer a Wilson depression, and that is something the traditional inversion method has no way of doing.

However, we advise care when training the neural networks on the atmospheres resulting from simulations. While the MHD simulations have, without doubt, reached a high degree of realism one should not forget that neural networks are quite bad at extrapolating outside of the domain of the training set, which  means that we can view the CNNs as extremely fast and clever interpolation methods. These methods will probably fail when faced with a phenomenon simulations have not predicted. In this publication we take  another approach to neural network inversions. We train the networks on data inverted by a conventional method (i.e., $\chi^2$ minimization) that has been chosen as a representative subset of the total data set to be interpreted. The network is then applied to the full data set. To verify this neural network inversion, we use the output of the network (i.e., the inferred physical parameters) to calculate ``predicted'' Stokes spectra. Finally we compare both the fit quality and the inferred parameters versus the ones of the actual inversion code, applied to the full data set. This allows us to  address the goodness of fit of the neural network inversions and  to discuss the difference in the inferred parameters with respect to the standard inversion procedure. We refer to this approach as mimicking the inversions since the neural network is trained on the inversion results and is basically trying to predict what the inversion method would infer if applied to the data. As most   machine learning approaches, this one also requires great care when selecting the training set. Strictly speaking, there is no guarantee that the required space of parameters is sampled by the training set. Therefore, we test three different approaches for selection of the training data set from the observations.

\section{Neural network setup and  training}

\subsection{Forward problem}

The generative model that describes spectral line formation can schematically be represented as
\begin{equation}
\vec{I}(\lambda) = \mathcal{F}\left [T(z),p(z),v_{\rm turb}(z), v_{\rm los}(z), \vec{B}(z) \right ]
\label{PRTE}
,\end{equation}
where $\vec{I}$ is the wavelength dependent emergent specific Stokes vector, and functional $\mathcal{F}$ involves all the radiative transfer processes and underlying physics and acts on relevant, depth-dependent physical quantities. Polarized spectral line formation is extensively covered in the literature and some of the excellent references are \citet{SAbook2014} for stellar atmosphere and spectral line physics, \citet{dtibook} for the Zeeman effect and spectropolarimetric diagnostics, and \citet{LL04} for details of spectral line polarization and specifically the Hanle effect. To illustrate the complexity involved in spectropolarimetric inversions, and for the completeness of this paper we briefly outline spectral line formation mechanism here.

Emergent polarized intensity is the solution of the polarized radiative transfer equation,
\begin{equation}
\frac{d\vec{I}(\tau)}{d\tau} = \hat{K}(\tau)\left ( \vec{I}(\tau) - \vec{S}(\tau) \right )
,\end{equation}
where $\hat{K}$ and $\vec{S}$ are the absorption matrix and the polarized Source function, respectively, and generally   depend on  all the atmospheric parameters. The parameter $\tau$ is the optical depth at the reference wavelength (usually at $5000\,\rm{\AA}$). This dependence involves various physical processes, the most important ones being ionization, excitation, natural line broadening, collisional and thermal line broadening, Zeeman splitting, and Zeeman polarization (selective absorption and magneto-optical effects). In general we should also add scattering line polarization and the Hanle effect to the list, but since there are no established inversion methods that involve stratified atmospheres and that account for these effects we defer them to future work.

The way the physical conditions in the atmosphere influence the absorption matrix and the polarized source function cannot always be expressed in an analytical way. For example, to obtain the number density of a given ion, we have to solve a   nonlinear system of equations involving the number densities of all the relevant ions, electrons, and molecules. Additionally, if photoionization and photoexcitation processes are not in equilibrium with the matter (i.e., if we are not in LTE), radiation must be included in the equations, making the problem much more complicated. Therefore, even the forward problem is computationally demanding.

\subsection{Inverse problem}

The solution of  Eq.\,\ref{PRTE} is given, in its integral form, as \citep[see, e.g.,][]{LL04, dtibook}
\begin{equation}
\vec{I}(\lambda)^{+} = \int_0^{\infty} \mathcal{O}_{\lambda}(\tau) \vec{S}_{\lambda}(\tau) d\tau 
,\end{equation}
where $\vec{I}^+$ is emergent Stokes spectrum to be compared and $\mathcal{O}_{\lambda}(\tau)$ is the so-called evolution operator at considered wavelength $\lambda$. In general, $\mathcal{O}_\lambda(\tau)$ and $\vec{S}(\tau)$ depend on all the atmospheric parameters. To infer the atmospheric parameters, given the emergent Stokes vector, we need to solve this integral equation. Solving integral equations is known as inversion, and hence the procedure of inferring atmospheric parameters from the Stokes spectrum is called spectropolarimetric inversion. We now briefly review the inversion procedure. 

The atmosphere and the dependence of all the quantities is discretized, typically using $ND = 50-100$ depth (height) points. This would make emergent intensity depend on the atmospheric parameters in the following way:
\begin{equation}
\vec{I}(\lambda) = f(T_1,T_2...T_{\rm ND},B_1,B_2...B_{\rm ND}...) = f(\vec{p}).
\end{equation}
We note  that the function that originally acted on parameter distributions is now replaced by a function that acts on all the discretized values of the atmospheric parameters. This is a function of   many variables (several hundred), and fitting it to an observed Stokes spectrum is a severely ill-posed problem. To remedy this the atmosphere is usually simplified using nodes. Nodes are pre-chosen depth points where the physical quantities are free to vary, while the values on the rest of the grid are interpolated (or extrapolated). This means that by knowing the transformation
\begin{equation}
\vec{\Theta} \rightarrow (\vec{p})
\label{par_to_nodes}
,\end{equation}
where $\vec{\Theta}$ is the vector containing values at all the nodes, we can define function $\vec{I}_{\lambda} = f(\vec{\Theta})$, which acts on the space of nodes. For example, if the atmosphere is  parameterized using five nodes in temperature, three in microturbulent and line-of-sight velocity, and one in magnetic field vector, the dependence would look like this:
\begin{equation}
\vec{I}_{\lambda} = f(T_1,T_2,T_3,T_4,T_5,\zeta_1,\zeta_2,\zeta_3,v_1,v_2,v_3,B,\theta_B,\phi_B) = f(\vec{\Theta}).
\end{equation}
Here $\zeta$ denotes turbulent velocity, $v$ line of sight velocity, and $\theta_B$ and $\phi_B$ are the inclination and azimuth of the magnetic field vector (i.e.,  our observable is a function of 14 parameters). Spectropolarimetric inversion is the process of finding the most probable parameters given the observations:
\begin{equation}
p(\vec{\Theta}|\vec{I}_\lambda^{\rm obs}) = \max.
\end{equation}
If all the values of all the parameters are equally probable {a priori}, this is the same as finding the parameters that minimize the $\chi^2$ merit function:
\begin{equation}
\chi^2(\vec{\Theta}) = \sum_{s,l} \frac{I_{s,l}^{\rm obs}-f_s(\vec{\Theta},\lambda_l)}{\sigma_{s,l}^2}.
\end{equation}
In practice, this minimum is usually found using a gradient-based technique. For more details on node-based spectropolarimetric inversions, see a recent paper by \citet{SNAPI} or a thorough review by \citet{LRSP}. We note that different implementations exist. For example, the SIR code \citep{SIR} parameterizes the corrections to the atmosphere using nodes instead of parameterizing the atmosphere itself, but the concept is similar to the one we described.

To summarize, spectropolarimetric inversion is an inference method that finds maximum likelihood solution (Bayesian approaches are, in principle, also possible)  for the observed polarized spectrum, given the model atmosphere parameterized using nodes.

\subsection{Neural network setup}

Neural networks can be understood as pipelines. Input data  enters at one end, and the neural network applies a series of simple transformations on it to produce the output. The transformations, in general, change   the dimensionality of data as well as the actual numerical values. Training the network   essentially means tuning   the coefficients of these simple transformations in order to match inputs and outputs of the given training set as closely as possible. After (and during) the training, the network is usually verified on the so-called validation set. This is a subset of the training data that has been set aside in order to check how well the network performs on a set of unseen data. Essentially, the network approximates the mapping from the input (polarized spectra) to the output (model parameters, in this case nodes). The first approaches to neural network inversions \citep[e.g.,][]{Carroll2001}, used densely connected layers. In this context, a layer is one operation acting on the input. For example, one dense layer relates input and output via matrix multiplication:
\begin{equation}
\vec{y} = \hat{M}\vec{I} + \vec{k}.
\end{equation}
These are usually followed by a nonlinear function so the output is
\begin{equation}
\vec{p} = f(\vec{y}) = f(\hat{M} + \vec{k}).
\end{equation}
That is, we first apply a linear operation on the input (in this case spectra), and then act on it with an activation function $f$, which allows the network to learn nonlinear mappings. Training the network means basically supplying a large enough training set of different $\vec{p}$ and $\vec{I}$ vectors so that elements of $\hat{M}$ and 
$\vec{k}$, and parameters of $f$ can be well constrained. This is, essentially, a regression to find best fitting $\hat{M}$, $\vec{k}$, and $f$. It is possible, in principle, to stack several densely connected layers to make a more complicated neural network, which would be capable of tackling more complex mappings, but the training would be more computationally intensive as well. 

With the advent of deep learning \citep{CNNref} we are witnessing new approaches to constructing layers. The most important ones are the convolutional layers that, as the name implies, apply convolutions to the input data. In astrophysics convolution commonly implies some type of smearing; however,  these convolutional kernels can play various roles (e.g., they can take shape of edge detection filter, etc.), and they seem to be particularly suitable for feature detection (e.g., recognizing objects in images). These layers are usually accompanied by the so-called pooling layers, which essentially bin the data together, thus reducing the dimensionality of the vectors involved. Networks with more than one layer are known as deep. Combining multiple convolutional, pooling, and densely connected layers results in a deep convolutional neural network. These architectures recently brought a huge boost to the field of machine learning and spawned a new direction (known as deep learning) enabling us to solve classification and regression problems with high accuracy and a huge increase in speed.

Deep convolutional neural networks have recently found their way into solar physics and the first applications are extremely promising. \citet{DeepVel} used them to automatically infer horizontal velocities in the solar atmosphere. \citet{CNNMOMFBD} showed that they can be trained on the examples of deconvolved data to allow for real-time multi-object multi-frame blind deconvolution (MOMFBD). \citet{Carlos_HMI_DL} used them to deconvolve and super-resolve HMI observations (this work also contains a very comprehensive introduction to deep convolutional networks). Finally, a recent paper by \citet{CNN_INV_AA} showed how to do extremely fast inversion of HINODE data using a convolutional neural network. Authors trained the network on the synthetic Stokes images of the solar surface, at various wavelengths, and related these images to the parameter maps at different optical depths. That is, their network was trained to learn the mapping: 
\begin{equation}
\vec{I}(x,y,\lambda) \rightarrow \vec{p}(x,y,\tau)
.\end{equation}
Here $\vec{p}$ is a vector of parameters (i.e., $\vec{p}=(T,p,v_{micro},v_{los},\vec{B}...)$) and $\tau$ is a coarsely sampled grid of optical depths. The network is trained on patches of the synthetic images that are result of MHD simulations, and various wavelengths and Stokes parameters are treated as channels (e.g., similarly to colors in the case of image processing). This network is convolutional in spatial domain, which allows easy generalization to arbitrary image sizes. Additionally, the training data was convolved with a known HINODE spatial point spread function, so the neural network automatically deconvolves the data.

In this work, we opted for a different path. We wanted a simpler and more data-driven approach. Specifically, we trained a network that can perform exactly the same task as an inversion code. That is, given the observed Stokes spectrum, the network estimates the most probable values of the pre-chosen model parameters (i.e., the nodes). The problem is then how to choose a good training set. Since we want a network that mimics what spectropolarimetric inversion is doing, we trained the network on a set of already inverted spectra. This approach, both in spirit and in the network architecture, is similar to the work of \citet{QSOCNN}, who use the network to identify and characterize the damped Lyman\,$\alpha$ systems from the observations of quasar spectra. The network is convolutional in wavelength and it performs both classification (to discriminate between different classes of objects) and estimation (to estimate redshift and H\,I column density). In addition, the authors use a set of objects whose properties are already inferred by an independent method as a training set. Another work that seems similar to ours is a recent paper by \citet{IRISCNN}, where deep learning is used to infer atmospheric structure from observations by the space-borne solar imaging spectrograph IRIS. Their work, however, lacks the detailed description of CNN architecture so we cannot compare it to ours in detail.

\begin{figure*}
\includegraphics[width=1.0\textwidth]{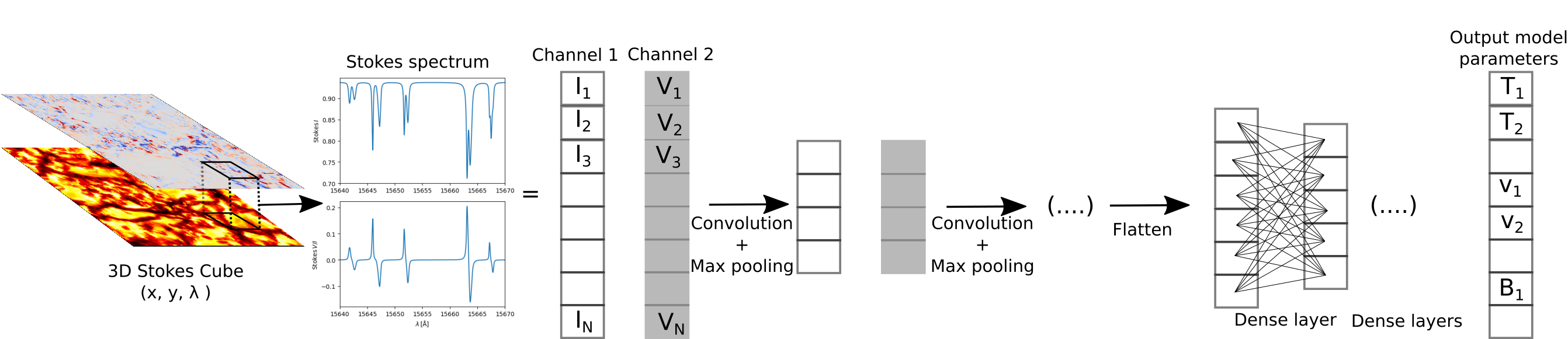}
\caption{Outline of the neural network architecture used in this paper. In principle, the number of convolutional and dense layers, as well as the width of the convolutional kernel, can vary depending on the data and the application.}
\label{CNN_arch}
\end{figure*}

We use a simple convolutional neural network that consists of several convolutional layers, each of them followed by a pooling layer. Next we use several densely connected layers, and then the output. The actual number of layers can vary depending on the type of data and the instrument used to obtain it. Different Stokes vectors are treated as different channels in the data. Convolution in wavelength basically means that we are interested in spectral features (i.e., lines) and the information they are carrying. Successive application of convolutional and pooling layers allow us to asses wavelength variations at different scales. In Fig.\,\ref{CNN_arch} we show the architecture of the neural network used in the example in Section \ref{sect:ex1}. That specific architecture has three convolutional layers, each followed by a pooling layer, and three dense layers (see the example below for more details).

It should be noted that this architecture implies that we treat each pixel separately. We relate an $N_s \times N_\lambda$ vector, where $N_s$ is the number of Stokes components taken into account and $N_\lambda$ is the number of wavelength samples, to the $N_P$ vector, where $N_P$ is the total number of nodes at each pixel. This inevitably brings some slowdown compared to the approach of \citet{CNN_INV_AA}, but the network is still many orders of magnitude faster than the state-of-the-art inversion codes. 

\subsubsection{Choosing a training set for the network}
\label{section:training_set}

Schematically,  a convolutional neural network   learns the mapping $\mathcal{M}$,
\begin{equation}
\vec{\Theta} = \mathcal{M}[\vec{I}_\lambda,\vec{k}]
,\end{equation}
by fitting the network parameters so the difference between $\vec{\Theta}^{\rm predicted}$  and $\vec{\Theta}^{\rm training}$ is minimal. To get the trained network which will perform well on various unseen data, we want to encompass all the various shapes of Stokes spectra and corresponding models (node values) in the training set. In the context of the solar atmosphere it means that if we want a neural network that reliably extracts physical parameters from an observed region, our training set should encompass all the characteristic spectra found there. For example, if we are observing a quiet Sun, the training set should involve spectra of granules, intergranular lanes, both in network and internetwork, and eventual bright points. If we are focusing on an active region, then, in addition to quiet Sun spectra surrounding the region, we should include sunspot umbra and penumbra, light bridges, and plage in the training set.

One way to select the training set is to do it ``by eye,'' in other words  to simply select a physical subregion of the observed region that encompasses all the various solar features. We tried this approach, and while it yields very good results, for the examples below we   tested two additional ways of selecting data for the training set:

\begin{enumerate}
  \item Random selection: From the whole field of view, we select $2\times10^4$ pixels at random. The assumption is that this is a large enough sample for each of the possible features in the data set to be well represented.
  \item K-means clustering: We first classify pixels in five subgroups using K-means clustering technique and then select, at random, $25\%$ of the pixels from each of the groups, but not more than $4\times10^3$ from each group. This again brings us to $\mathcal{O}(10^4)$ (maximum of $2\times10^4$) spectra for the training. We note that we tested agglomerates in place of K-means clustering because the spectra can in principle have a large number of dependent dimensions (i.e., wavelengths), and in that case agglomerate clustering is more appropriate. Unexpectedly, the training set selected by agglomerate clustering gave worse results than the other three sets described above. We ignored the set obtained with agglomerate clustering for now as the poor performance might be a consequence of our using it inappropriately.
\end{enumerate}

After selecting the training set we perform a Levenberg--Marquard-based inversion on the training set to obtain the most probable values of atmospheric parameters. These parameters are then normalized so their values all fall in the range $[0,1]$. This is important as the physical parameters span different magnitudes in  CGS units. The magnetic field is also split into a horizontal and vertical component, similar to what is done in \citet{CNN_INV_AA}. Additionally, the input (Stokes parameters) are normalized with respect to the quiet Sun continuum, as is commonly done in the case of spectropolarimetic inversions \citep[e.g.,][]{dtibook}.

We then train the neural network using the stochastic optimizer ADAM \citep{adam}, until the convergence, which is  around 80-100 epochs for all of the sets. After the training, neural network parameters are saved and can be used for the interpretation of the full data set or other data sets, provided the spectra have the same wavelength sampling. We note that the network will return transformed parameters in the range $[0,1]$ so these need to be transformed back to the appropriate physical units. 

The network is set up and trained using Keras \citep{KERAS} and Tensorflow \citep{TENSORFLOW} frameworks in the python programming language. All the network layers, training, minimizers, and validations are implemented in this packages so the user only has to worry about the setup of the network, the choice of the layers, and the selection and the normalization of the training set.

\section{Application to the synthetic data}
\label{sect:ex1}

We start with a relatively simple example of the synthetic data, from atmospheres resulting from an MHD simulation run. Working with synthetic data is advantageous because we can directly compare inferred quantities with the ground truth. We note that we can only expect the convolutional neural network to perform as well as the inversion method itself. Synthesizing spectra from MHD simulations and inverting them to test the inference method performance, biases, and accuracy is a common practice; however it is  not extensively documented \citep[but see, e.g., ][]{MEtests, IRFe}.

In this example we start from an MHD cube representing quiet Sun region with mean vertical magnetic field of 50\,gauss. We synthesize the five infrared Fe\,I lines around 15600\,\AA, which are well known for their high magnetic sensitivity and low formation height, making them ideal for spectropolarimetric diagnostics of photospheric layers. These are the same lines and the cube that are used in \citet{IRFe}, and we refer the interested reader to that paper for the details on the lines, synthesis, and the inversion. We synthesized the spectra in the region from \rm{15640\,\AA} to \rm{15670\,\AA} with $60\,\rm{m\AA}$ sampling in the direction $\mu=\cos\theta=1$ (where $\theta$ is the heliocentric angle). We then convolved the data in wavelength to simulate finite spectral resolution of the instrument (in this case $\approx 10^5$), and added wavelength-dependent photon noise assuming S/N$\approx 1000$ in the quiet Sun continuum. We then inverted Stokes $I$ and $V$ of the whole set of observations ($288\times288$ spectra), using the SNAPI code \citep{SNAPI}, assuming a model described in Table \ref{table:nodes}. The result of this inversion are the values of the physical parameters at the nodes for the whole simulated field of view and the corresponding atmospheres obtained from the nodes using assumed relationship from Eq.\,\ref{par_to_nodes}. From now on, by model parameters we mean values of physical parameters at the node locations.

\begin{table}[]
   \caption{Choice of the nodes used for inversion of the synthetic data set}
       \centering
       \begin{tabular}{c c}
        Parameter & Node positions [$\log\tau_{5000}$]\\ \hline   
        Temperature   &  -3.4, -2.0, -0.8, 0.0, 0.5 \\
        LOS velocity & -2.5, -1.5, -0.5, 0.5 \\
        Magnetic field strength & -1.5, 0.3 \\
        Microturbulent velocity & const \\
        Magnetic field inclination & const
       \end{tabular}
    
       \label{table:nodes}
\end{table}

We then select a training set for our neural network from these $288 \times 288$ pixels, using each of the methods we described in section \ref{section:training_set}: selecting a subregion by eye (we chose central $144 \times 144$ pixels), random sampling of pixels, and K-means clustering followed by random sampling. We train the network on each of the data sets and compare the final validation errors between the three approaches.  The validation error tells us how well the network performs on the  so-called validation set, a randomly selected subset of the training data. We note that for each pair $(\vec{I}_\lambda,\vec{\Theta})$ we relate the model parameters to the calculated (i.e., fitted) spectra and not the original observed spectrum. This ensures that the input and output are related by our assumed physical model.

We then apply the network on the whole set of the synthetic Stokes profiles and make the following comparisons:
\begin{itemize}
  \item Mean and median $\chi^2_{\rm reduced}$ for each of the CNN inversions. We obtain these by feeding the inferred models back in the forward calculation and comparing the resulting spectra with the original synthetic spectra that was used for inference.
  \item Agreement between the inferred parameters and the referent solution, i.e., the parameter values obtained by the standard inversion method, in this particular case, the SNAPI spectropolarimetric inversion code. 
\end{itemize}

For this example we use a convolutional network that has three convolutional layers with convolutional filters with width of 7, 5, and 3, respectively, each   followed by a pooling layer of width 2. These layers treat spectral lines as features. All the convolutional layers are activated using the rectified linear unit (ReLU) activation. Finally, we have three successive densely connected layers with decreasing dimensionality, the last one having the same number of neurons as the number of output nodes. Between the first and second dense layers we include a dropout layer that is generally used to decrease overfitting and make neural network better at generalizing. The dense layers have sigmoid activation with the exception of the last one, which has linear activation. We found this combination of activations to work the best, although differences with just using ReLU everywhere are small. This network is technically ``deep'' since it involves more than one hidden layer, but it is significantly simpler then the ones used by \citet{CNN_INV_AA} or \citet{RADYNINV}, among others. It is very similar to the one used by \citet{QSOCNN}. For this example we interpret only Stokes $I$ and $V$, and thus do not try to infer the magnetic field azimuth.

%Finally, we also compare inferred quantities versus original values from the MHD cube, at appropriate depths, similar to what was done in \citet{IRFe}. {\color{red} Do I do this at all? Sounds like just repeating stuff from Fe paper...}

\begin{table}[]
   \caption{Performance of the neural network using the three different data subsets. $\chi^2$ values are obtained by substituting the inferred model parameters   in the forward model, and comparing the obtained spectra with the observed ones.}
       \centering
       \begin{tabular}{c | c c c }
        Measure & Set 0 & Set 1 & Set 2 \\ \hline   
        Validation error & 0.0013 & 0.0013 & 0.0020 \\
        Set size & 20736 & 20000 & 8631 \\
        $<\chi^2_{\rm reduced}>$ & 62.73 & 66.7 & 51.0 \\
        $\rm{median\,\chi^2}_{\rm reduced}$ & 33.3 & 34.3 & 33.3 \\
       \end{tabular}
    
       \label{table:cnn_results1}
\end{table}

Table \ref{table:cnn_results1} contains the results obtained by training the network on three different subsets and then applying it to the whole set of synthetic spectra. The validation error is the smallest in the case of the training region chosen by eye. We explain this with the fact that atmospheres that are physically close to each other in the simulation have similar properties and then the validation set is the most similar to the training set. The validation error is the greatest for  training set 2 (i.e., the one chosen by K-means clustering). This can be explained by the fact that this training set was the smallest and hence the training quality in general was worse. 

\begin{figure*}
\includegraphics[width=\textwidth]{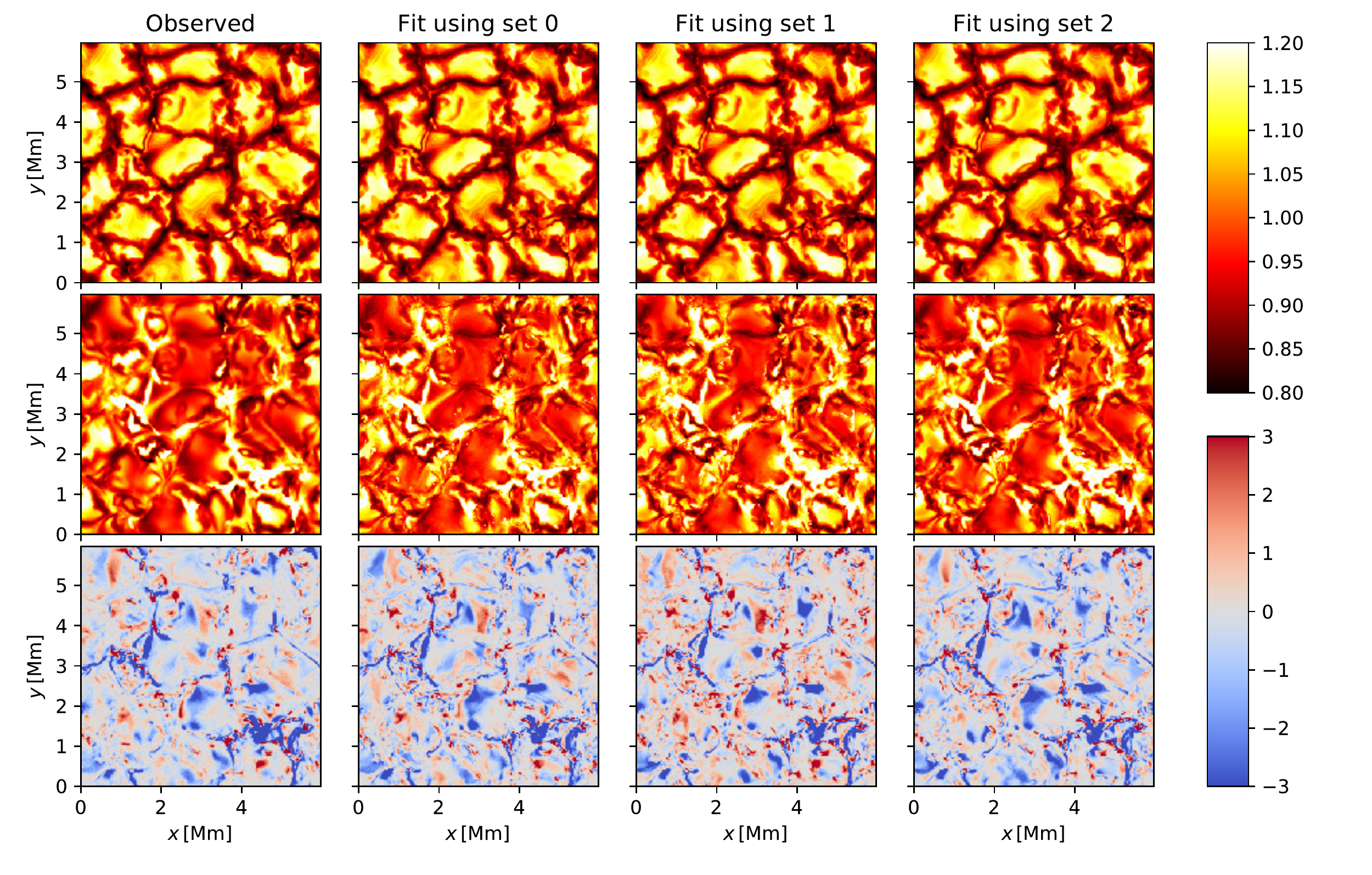}
\caption{Comparison between the spectra calculated from the models inferred by the CNN and the original synthetic observations. Top: Continuum; middle: Core of $15662\,\rm{\AA}$ line, bottom: Circular polarization (Stokes V)  in the near wing of $15662\,\rm{\AA}$ line.}
\label{Fig1:inversions}
\end{figure*}

Interestingly, the fit is the best in the case of the training set 2, even though it is the smallest. The reason  is that K-clustering ensures a much-needed diversity of the training set and provides the best fits for the unseen data set. This means that this training set provides the network with the best generalizing power. In addition to calculating the $\chi^2$, we show the images at specific wavelengths for the synthetic observations and each of CNN inversions in Fig.\,\ref{Fig1:inversions}

Judging from the single wavelength, it would seem that the training set that reproduces the observed polarization the worst is the last one. However, calculating $\chi^2_{\rm reduced}$ for Stokes $V$ only yields mean values of 6.1, 5.5, and 5.3 for the three training sets, which is not a dramatic difference. Visually, neural network reproduces the observed intensity and polarization very well, excluding some small details in Stokes V. For the comparison, the $\chi^2_{\rm reduced}$ value we obtained with a standard inversion method (in this case using SNAPI inversion code) has mean of 49.3 and median of 18.6. This means that the average fit between the standard inversion and the neural network case is very similar, but the standard inversion provides better fit for the majority of the pixels. This means that the neural network is good at roughly estimating the location of the minimum, while the standard inversion code is better at pinpointing the exact minimum. The question is, of course, how close to the minimum  the solution obtained by the neural network is.

To answer this question and further evaluate the performance of the neural network we directly compare values of parameters at the nodes, inferred by using each of the training sets versus the values inferred by SNAPI. In addition to visual comparison, we show the median of the difference and the 90\% confidence interval (i.e., the 5th and 95th percentile). This tells us whether CNN inversion is shifted or skewed with respect to the referent inversion. We focus on the depths these lines are the most sensitive to, as shown by previous studies \citep[e.g.,][]{IRFe}. Specifically, we look at the temperature at $\log\tau=0$ (photosphere), $\log\tau=-0.8$ (approximately inverse granulation layer), line-of-sight velocity at $\log\tau=-0.5$ (mid - photosphere), and magnetic field at $\log\tau=0.3$ (photosphere) and $\log\tau=-1.5$ (upper photosphere). 

\begin{table}[]
   \caption{Comparison between the model parameters of interest obtained with the SNAPI inversion code and with the convolutional neural network trained on the three different subsets. Shown are the median of the difference between two respective results, and the upper and lower uncertainty that define the 90\% confidence interval.}
       \centering
       \begin{tabular}{c | c c c }
        Quantity & Set 0  & Set 1  & Set 2  \\[5pt] \hline   
        $T(\log\tau=-0.8)\,\rm{[K]}$ & $-24_{-122}^{+115}$ & $-41_{-106}^{+121}$ & $-26_{-113}^{+122}$ \\[5pt]
        $T(\log\tau=0)\,\rm{[K]}$ & $-11_{-90}^{+94}$ & $-13_{-87}^{+80}$ & $-14_{-86}^{+76}$ \\[5pt] 
        $v_{\rm los}(\log\tau=-0.5)\,\rm{[km/s]}$ & $0.06_{-0.55}^{+0.49}$ & $0.07_{-0.57}^{+0.45}$ & $0.07_{-0.54}^{+0.49}$ \\[5pt]
        $B(\log\tau=-1.5)\,\rm{[gauss]}$ & $-2_{-54}^{+56}$ & $2_{-58}^{+41}$ & $-5_{-62}^{+47}$  \\[5pt]
        $B(\log\tau=0.3)\,\rm{[gauss]}$ & $5_{-65}^{+77}$ & $10_{-68}^{+77}$ & $6_{-60}^{+76}$ \\[5pt]
       \end{tabular}
    
       \label{table:cnn_results2}
\end{table}

\begin{figure*}
\begin{centering}
\includegraphics[width=1.0\textwidth]{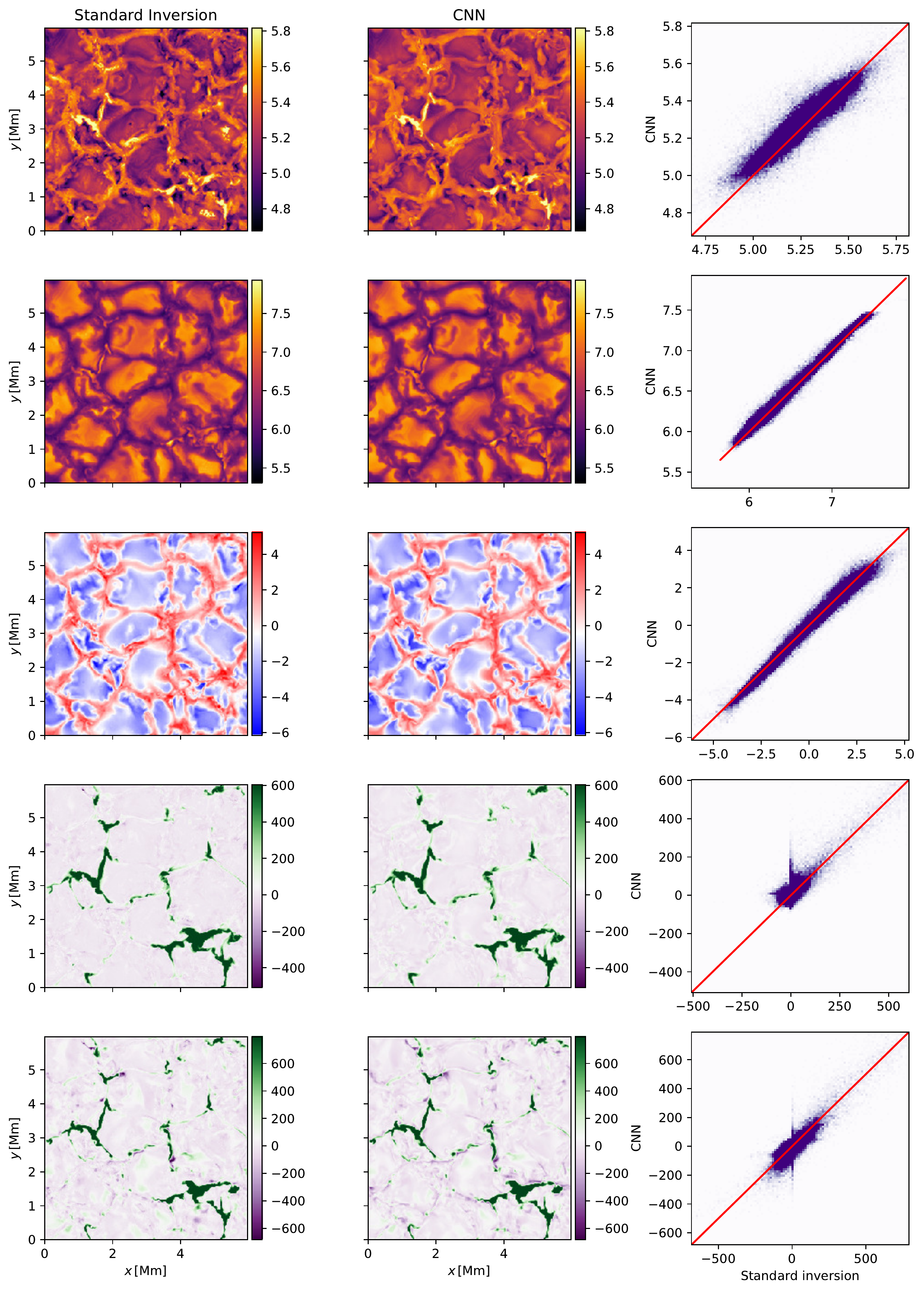}
\caption{Comparison between the node values inferred by a standard, maximum likelihood-based, inversion code, and the results of the convolutional neural network, using set 2 as the training set. First two rows: Spatial parameter distributions. Third row: Scatter plot between the two. From top to bottom: temperature at $\log\tau=-0.8$ and $\log\tau=0$; line-of-sight velocity at $\log\tau=-0.5$; line-of-sight magnetic field at $\log\tau=-1.5$ and $\log\tau=0.3$. Units are kK, km/s, and gauss, for temperature, velocity, and magnetic field, respectively.}
\label{Fig2:parameters}
\end{centering}
\end{figure*}

The comparison between the inferred parameters is summarized in  Table \ref{table:cnn_results2}. We also show the the maps of the parameters of interest inferred by  standard inversion (SNAPI) and the neural network (Fig.\,\ref{Fig2:parameters}) using training set 2. The agreement of all the parameters is very good, where systematic offsets are much smaller than the upper and lower uncertainties.  The offsets and the uncertainties are lower than typical differences between the original MHD cube and the values retrieved by an inversion  \citep[see, e.g.,][]{IRFe}. Not surprisingly, the most reliable inference is for the photospheric temperature and the velocity, the quantities that regular inversion codes also routinely retrieve very reliably. The magnetic field between the two methods also agrees very well, but  artifact-like disagreements around low magnetic field values should be noted. We recall that we have added photon noise to our synthetic data, and hence the very low values of the magnetic field cannot be retrieved. The magnetic field in the upper layers is retrieved more reliably, which is in  line with the conclusions in the Appendix of \citet{IRFe}. 

\begin{figure*}
\includegraphics[width=1.0\textwidth]{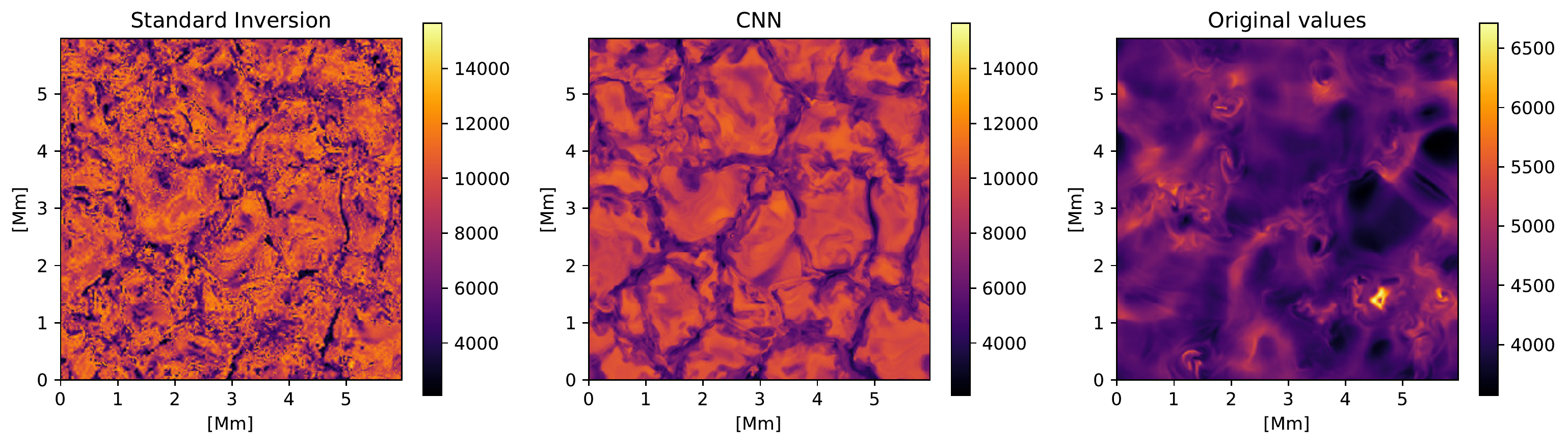}
\caption{Similar to \ref{Fig2:parameters}, but for  the temperature at the node at $\log\tau=-3.4$. The rightmost plot is the original atmosphere from the simulated atmosphere. There is  a different range for the third plot.}
\label{Fig3:parameters}
\end{figure*}

From all the analyzed parameters, the agreement is the worst for the temperature at the node at $\log\tau=-0.8$. We note that the temperature map inferred by the neural network looks similar to the one inferred by the standard inversion, but is noticeably more smooth. Our interpretation is the following: the upper layers of the atmosphere are typically probed by fewer spectral points. The standard inversion method is more susceptible to random noise and is, in a way, trying to fit that noise, hence producing a salt-and-pepper pattern in the parameter maps. The neural network is, through the convolutional layers, able to reduce the dimensionality of the Stokes profiles and relate that reduced basis to node values, hence decreasing the sensitivity of the inference to the photon noise. To illustrate this we show the temperature at the node placed at $\log\tau=-3.4$, a parameter that is very poorly constrained by the data. The spectra themselves are very weakly sensitive to this parameter, since lines are formed relatively deep, and thus the inferred value is very sensitive to the noise and is not an indicator of the actual temperature at $\log\tau=-3.4$ (this node basically only sets the value of the derivative for the deeper layers). The agreement between the two results is shown in Fig.\,\ref{Fig3:parameters}. The smoothness of the result obtained with CNN is obvious, but it does not make it more accurate or realistic. In particular, the temperature distribution evidently follows the granulation pattern, thus suggesting that temperature at this node is very degenerate (i.e. correlated) with the deeper ones. To illustrate this, we show the original MHD atmosphere in the rightmost plot of Fig.\,\ref{Fig3:parameters}. We note that although the parameter map is very smooth and looks well constrained, it is not indicative of the temperature of the actual atmosphere at all. Thus, when interpreting the results of neural networks, knowledge of the line formation properties is needed, just as in the case with standard inversions.

\subsection{Preserving the correlations between the parameters}

When analyzing observations of high spatial resolution it is common to find some correlation between the inferred parameters; for example, granules are hotter and have negative velocities, while intergranules are cooler, have positive velocities, and stronger magnetic fields. It is interesting to test whether the neural networks preserve these correlations or if they perhaps induce some additional biases. To test this we calculate the Pearson correlation coefficient between the products of different inferred parameters:
\begin{enumerate}
  \item Temperature at $\log\tau=0$ and $v_{\rm los}$ at $\log\tau=-0.5$ (i.e., photospheric temperature and velocity).
  \item Magnetic field at $\log\tau=0.3$ and $v_{\rm los}$ at $\log\tau=-0.5$ (photospheric magnetic field and velocity).
  %\item Magnetic field at $\log\tau=-1.5$ and $v_{\rm los}$ at $\log\tau=-1.5$, magnetic field and velocity at the upper photosphere.
\end{enumerate}

\begin{figure}
\includegraphics[width=0.5\textwidth]{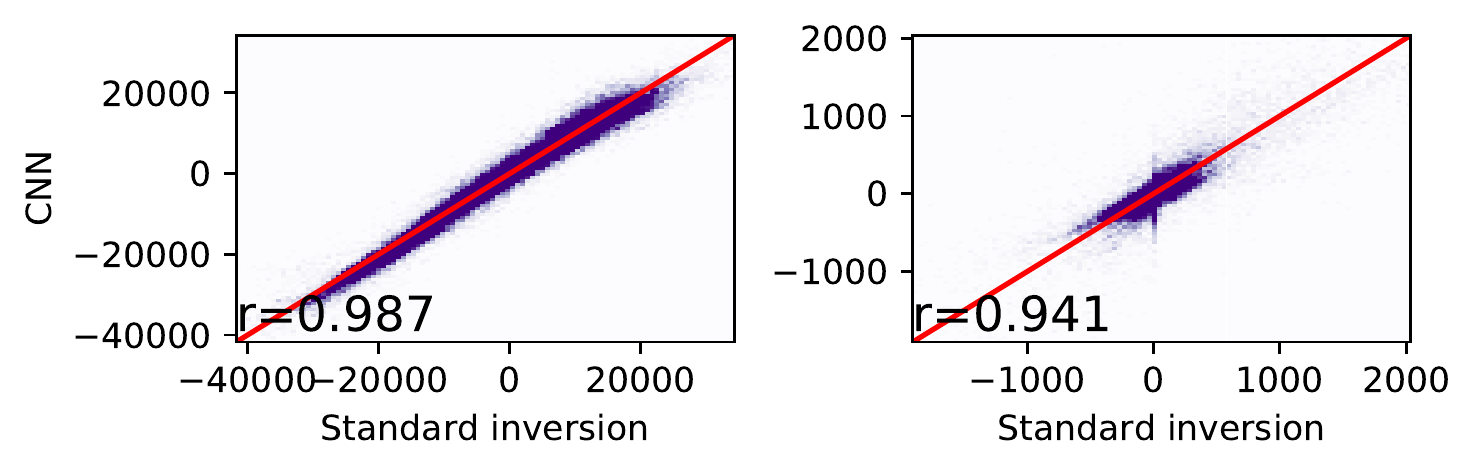}
\caption{Scatter plots and the correlation coefficients between the products of different node values. Left: $T$ at $\log\tau=0$ and $v_{\rm los}$ at $\log\tau=-0.5$; right: $B_{\rm los}$ at $\log\tau=-1.5$ and $v_{\rm los}$ at $\log\tau=-1.5$.}
\label{Fig4:corr_of_corr}
\end{figure}

Scatter plots between these parameters and the values of the Pearson correlation coefficient are given in Fig.\,\ref{Fig4:corr_of_corr}. There are no obvious biases (i.e., structures in the scatter plots). Although our testing sample here is very limited, this is a good indicator that neural network reproduces appropriate correlations between the parameters on the training interval. A true test would be the application of this neural network to  synthetic spectra from an MHD cube where the commonly encountered correlations are reversed or completely randomized (e.g., intergranular lanes have negative velocities). However, we note that our approach aims to train a neural network on a representative subset of the observed data, and not to map the inversion problem in general.
 
\section{Application to a different MHD run}
\label{sect:ex2}

To further test the neural network trained in Sect.\,\ref{sect:ex1}, we apply it to a different set of synthetic observations. We again use an atmosphere resulting from the MURAM code, but we chose a local dynamo run described in \citet{Dynamo_run_Rempel}. The difference with respect to the cube used for training is that there is no mean magnetic field imposed on the lower boundary and the spatial sampling is different. This should, in general, lead to a different stratification of the atmospheres. The whole simulated atmosphere has dimensions $1536\times1536\times480$. Since the simulation contains only the quiet Sun, we focus on a $400\times400\times120$ subset that contains several granules and vertically includes the regions relevant for the line formation.

\begin{figure*}
\includegraphics[width=\textwidth]{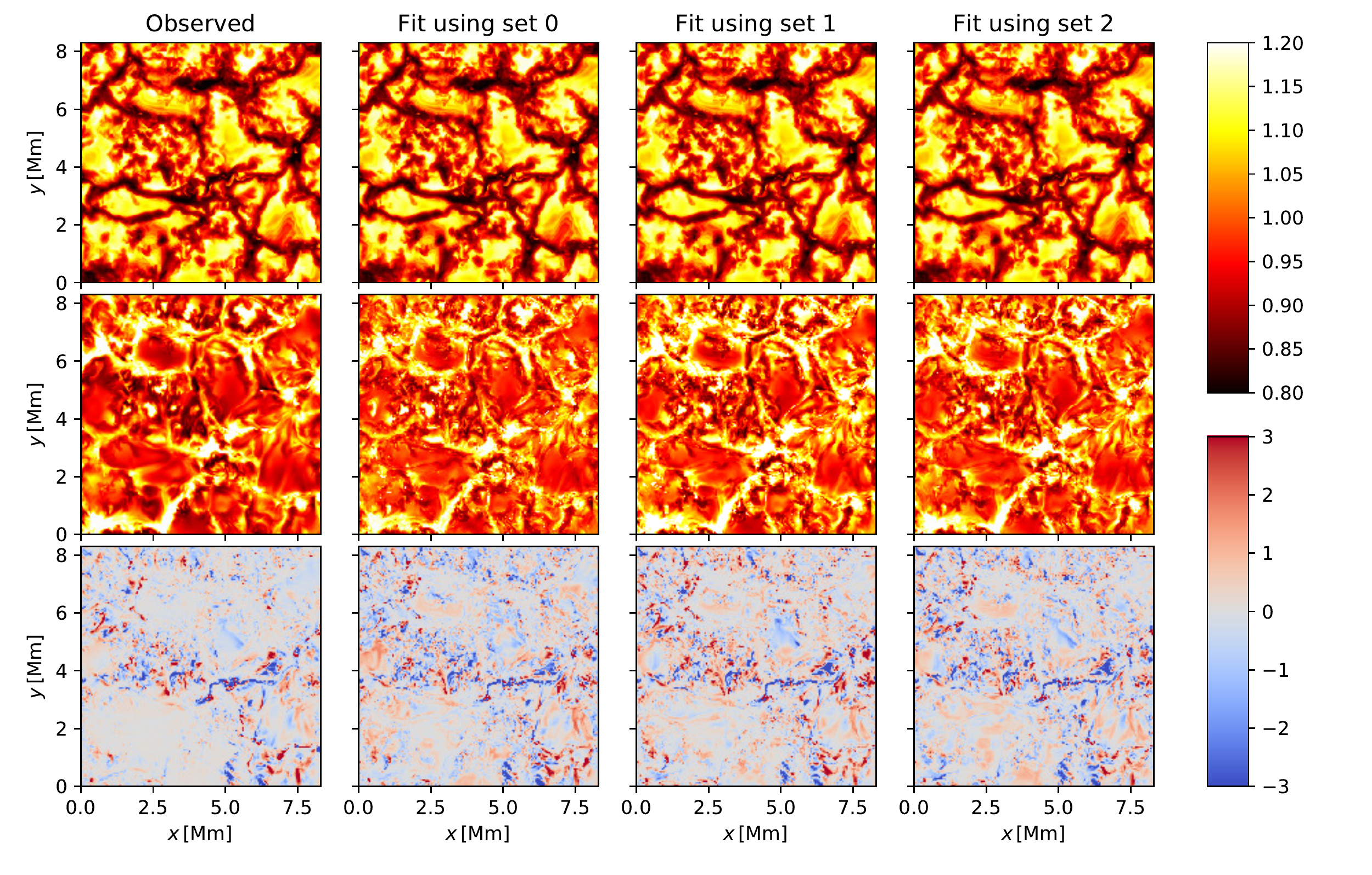}
\caption{Same as Fig.\,\ref{Fig1:inversions}, but for a set of synthetic data coming from a local dynamo simulation run.}
\label{Fig:inversions2}
\end{figure*}

We synthesize the lines in the same spectral window, with the same sampling and the  noise as in the previous example. We then apply the convolutional neural network trained on the three different sets described in Sect.\,\ref{sect:ex1}, but without any modifications. The fits are given in Fig.\,\ref{Fig:inversions2}. 

\begin{table}[]
   \caption{Same as Table\,\ref{table:cnn_results1}, but    already trained networks are applied  on the spectra synthesized from a local dynamo simulation.}
       \centering
       \begin{tabular}{c | c c c }
        Measure & Set 0 & Set 1 & Set 2 \\ \hline   
        $<\chi^2_{\rm reduced}>$ & 83.6 & 71.0 & 56.9 \\
        $\rm{median\,\chi^2}_{\rm reduced}$ & 39.4 & 28.0 & 25.9 \\
       \end{tabular}   
       \label{table:cnn_results3}
\end{table}

\begin{figure*}
\begin{centering}
\includegraphics[width=0.85\textwidth]{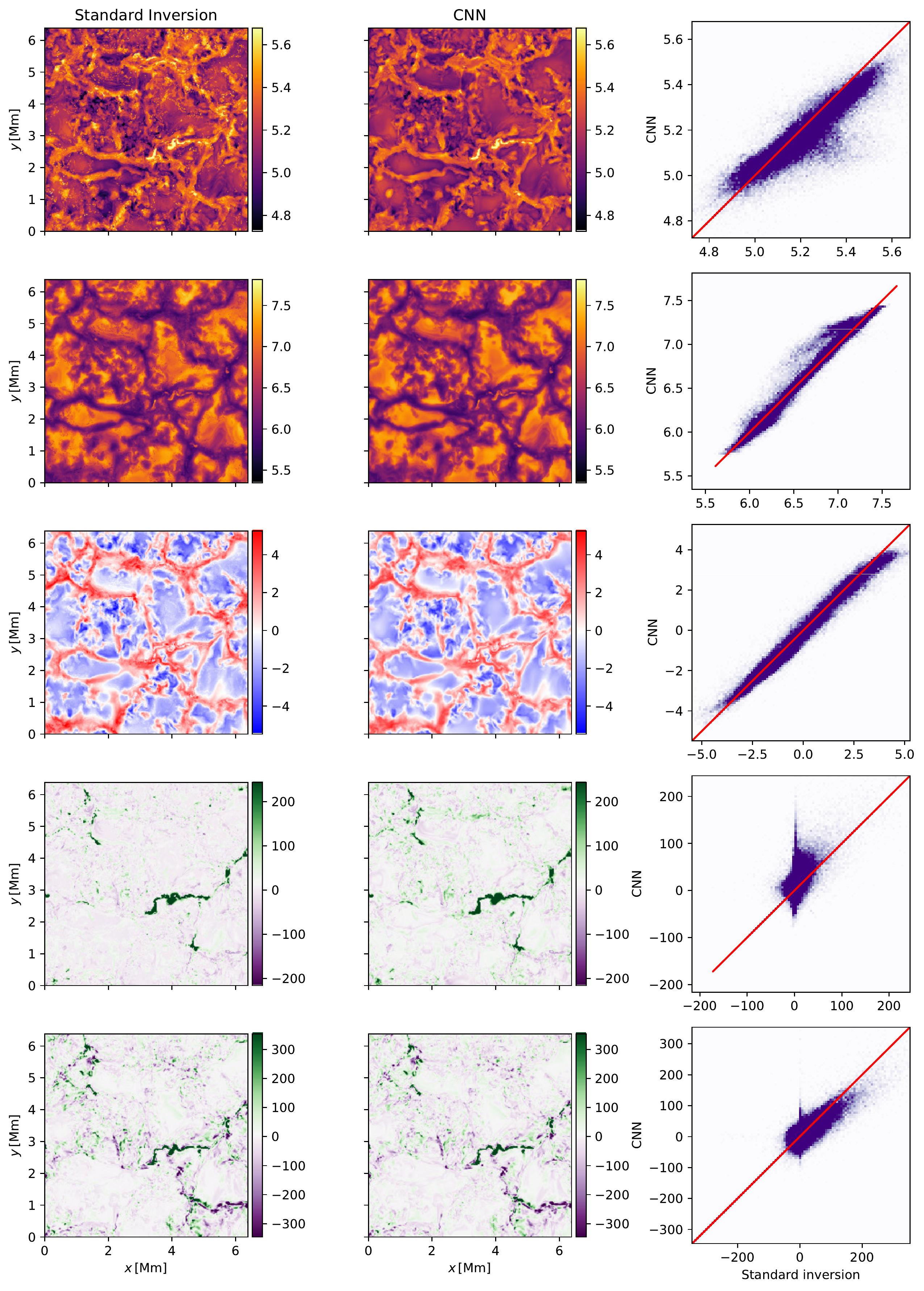}
\caption{Same as Fig.\,\ref{Fig2:parameters}, but for  a comparison of inversion code and the neural network on completely unseen synthetic data, from a different MHD atmosphere.}
\label{Fig:parameters3}
\end{centering}
\end{figure*}

Judging from Fig.\,\ref{Fig:inversions2}, the fits look good and there are no obvious artifacts or disagreements (we show only two wavelengths). To evaluate the agreement in more detail we calculated average and median $\chi^2_{\rm reduced}$ for each of the neural network inversions and listed them in Table\,\ref{table:cnn_results3}. Interestingly, the agreement is slightly better than in the previous case, although in the previous case we were using the network trained on the subset of the data to be interpreted. This shows that the network is doing a very good job at generalizing the inference to the data resulting from unseen simulations. However, the standard inversion applied to this data set, yielded even better fits, with mean $\chi^2_{\rm reduced}=29.6$ and median $\chi^2_{\rm reduced}=9.9$. Obviously, in this case, the standard inversion code found much better fitting atmospheres than the neural network. To asses this further, we plot the agreement between the relevant node values, in the same manner to the previous example, and calculate the median of the difference, as well as 90\% confidence interval (Table\,\ref{table:cnn_results4}). We show the comparison between the model parameters inferred using SNAPI and the neural network in Fig.\,\ref{Fig:parameters3}. These are the parameters inferred using training set 2 (as it produced the best fits), but the other two sets perform very similarly. 

\begin{table}[]
   \caption{Same as table\,\ref{table:cnn_results2} except we now compare inversion code and the neural network on completely unseen synthetic data, from a different MHD atmosphere.}
       \centering
       \begin{tabular}{c | c c c }
        Quantity & Set 0  & Set 1  & Set 2  \\[5pt] \hline   
        $T(\log\tau=-0.8)\,\rm{[K]}$ & $-12_{-106}^{+143}$ & $-36_{-89}^{+146}$ & $21_{55}^{+148}$ \\[5pt]
        $T(\log\tau=0)\,\rm{[K]}$ & $-44_{-110}^{+84}$ & $-23_{-109}^{+70}$ & $6_{-103}^{+56}$ \\[5pt] 
        $v_{\rm los}(\log\tau=-0.5)\,\rm{[km/s]}$ & $0.07_{-0.47}^{+0.53}$ & $0.18_{-0.40}^{+0.49}$ & $0.05_{-0.28}^{+0.50}$ \\[5pt]
        $B(\log\tau=-1.5)\,\rm{[gauss]}$ & $-2_{-49}^{+41}$ & $11_{-36}^{+39}$ & $-8_{-44}^{+11}$  \\[5pt]
        $B(\log\tau=0.3)\,\rm{[gauss]}$ & $11_{-55}^{+53}$ & $-27_{-49}^{+46}$ & $3_{-25}^{+40}$ \\[5pt]
       \end{tabular}
    
       \label{table:cnn_results4}
\end{table}

Judging from these simple statistical measures of the differences, the performance of the neural network on this synthetic data set is as good as in the previous example. We find this extremely important since the new simulated atmosphere is expected to have different stratification due to different boundary conditions. This is especially true for the magnetic field as in the case of the local dynamo run we are expecting more tangled and generally less smooth magnetic fields. However, it seems that a convolutional neural network is capable of inferring the magnetic fields that compare to the ones inferred by a standard inversion code. We note that this does not, by any means, imply that the inferred atmospheres represent the actual ones. Our tests simply show that training a neural network on the results of already performed inversions results in a network that generalizes well to other similar data.

\section{Conclusions}

In this paper we explored a simple convolutional network architecture with the goal of mimicking an inversion code. We tested it on a simulated Stokes cube containing spectra of five infrared Fe lines around 15600$\,\AA$, synthesized from a quiet Sun MURAM run with mean magnetic field of 50\,gauss. We selected a representative subset of the data, inverted it using the SNAPI code \citep{SNAPI}, and trained the network on the mapping $\vec{I}_{\lambda}\rightarrow \vec{\theta}$, where $\vec{\theta}$ is the vector of model parameters (node values). We tested three different methods of choosing the representative subset and found small differences. Selecting representative data by first doing a K-means clustering and then randomly selecting the data from different clusters yielded the smallest training set but the best fits, both in terms of $\chi^2$ and in terms of agreement with the results of the referent inversion code. The network generalized very well both to the original data set and to a data set produced by a different simulation run. Despite the differences in the simulation, the obtained $\chi^2$ values and the agreement with the results obtained with the standard inversion code were similarly good. 

The main advantage of the convolutional neural network in this case is speed. Inverting a $400\times400$ field of view with $400$ wavelength points at each pixel takes around ten seconds. For  comparison, the fastest inversion code known to the authors, SIR \citep{SIR}, needs several seconds for each spectrum. This is already an increase of a factor of $\approx10^5$. A common question we encountered in the discussions was the duration of training. A training on a single CPU from a training set of size $\approx 10^4$ spectra takes roughly an hour. It is possible that for a more general neural network (e.g., one involving lines with various sensitivity regions and/or different solar features) would take longer,  because of the larger training set and, possibly, the more complicated network. However, using GPUs would reduce training time immensely, thus making it unlikely that we will ever need more than a few hours to train the network. Once trained, the network can be applied to an arbitrary amount of data, obtaining results much faster than any standard inversion code. Actually, the slowest part of our calculation is the synthesis of the spectra from inferred models by a neural network ($\approx 0.1$\,s per pixel). We want to emphasize the CPU time saving provided by a neural network approach. We strongly believe that a detailed analysis of the spectra of NLTE lines obtained with high spatial and temporal resolution will be impossible with standard approaches, and that neural networks can provide the fastest, albeit approximate, diagnostics.

From these initial results we can propose a method for very fast inversion using a convolutional neural network:
\begin{itemize}
  \item Cluster the whole observed data set, making sure that there are enough clusters to cover all the different spectra shapes. This takes some trial and error, but it can be important for the selection of the appropriate training set.
  \item Randomly select a subset from each of the clusters,  and merge them in an aggregate training subset of size $10^4-10^5$.
  \item Invert the aggregate subset using the inversion code of choice. 
  \item Train the network using best fit profiles as the input and the best fit model parameters as the output. We use the best fit profiles as the input instead of the observed ones to avoid the systematics, for example caused by possible local minima and wrong fits. However, using the observed data helps offset other systematic effects such as fringes, telluric lines, etc. Choosing one or the other might depend on the specific spectral region and the instrument.
  \item Apply the trained network to the rest of data to infer the model parameters of the full data set (or other data sets done with the same observational setup).
  \item If desired, synthesize the predicted spectra from inferred model parameters to check for pixels with conspicuously bad fits, and analyze them separately.
\end{itemize}
It should be noted that this approach implicitly assumes that both the training data and the data to be interpreted are observed at the same heliocentric angle ($\mu$). If the observations are done with the same setup but at significantly different heliocentric angles the process above should be repeated. The same goes for the changes in the wavelength sampling, and possibly for drastically different exposures that result in a significantly different signal-to-noise ratio.

For instruments that observe at various heliocentric angles, but keep all other parameters exactly the same (e.g., HINODE/SP), this  limitation can be generalized somewhat. We can create a reference library of the models and calculate the Stokes spectra for a grid of heliocentric angles. The network then takes the heliocentric angle as an input, in addition to Stokes spectrum, and infers the model parameters (node values).

In our preliminary tests we found that this approach works well on spectral lines formed in different regimes (we tested on Fe\,I\,6300, Na\,I\,D, Ca\,II\,8542, Mg\,I\,b, etc.), real-life spectrograph data, and  filtergraph data (in this case it is not necessary to use convolutional layers). In the follow-up publication we will apply this approach to the full Stokes observations, and the training set obtained with a different inversion code.

The codes and the plotting routines used to for this publication are publicly available. \footnote{\href{https://github.com/ivanzmilic/deepinversion}{https://github.com/ivanzmilic/deepinversion}.}

\begin{acknowledgement}
We thank Andr\'{e}s Asensio Ramos, Christopher Osborne, Carlos Jose D\'{i}az Baso, Mark Rast and Mark Cheung on discussions and suggestions that led to this investigation. Comments by Mom\v{c}il Molnar, Shah Mohammad Badaouin and David Orozco Suarez greatly improved the manuscript. We also thank Tino Riethm\"{u}ller and Matthias Rempel for providing us with the MURAM simulations. We thank the anonymous referee for their comments that significantly improved the manuscript.

All the network training and inference has been done using Keras with Tensorflow package. All the plots were done using matplotlib python package.

This work has been supported by the Spanish Ministry of Economy and Competitiveness through projects ESP-2016-77548-C5-1-R and by Spanish Science Ministry ``Centro de Excelencia Severo Ochoa'' Program under grant SEV-2017-0709 and project RTI2018-096886-B-C51.” 
\end{acknowledgement}

\bibliography{inversion}

%-----------------------------------------------------------

\end{document}